\documentclass[epj]{svjour}

\usepackage{amsmath}
\usepackage{amssymb}
\usepackage{verbatim}
\newcommand{\be}[1]{\begin{equation} #1 \end{equation}}
\usepackage{graphicx}
\usepackage{fancyhdr}
\def\makeheadbox{{
 }}
\def\mytitle{My title} 
\def\myauthors{My name}  
\def\mytype{My type of session}
\def\mysession{My session}

\def\mytitle{O'Raifeartaigh models with spontaneous R-symmetry breaking} 
\def\myauthors{Luca Ferretti}    
\def\mytype{Contributed Talk}    
\def\mysession{Theoretical Models}

\begin{document}
\title{O'Raifeartaigh models with spontaneous R-symmetry breaking}
\author{Luca Ferretti
\thanks{\emph{Email:} ferretti@sissa.it}%
}                     
%
%
\institute{SISSA/ISAS and INFN, Trieste, Italy
}
%
\date{}
\abstract{O'Raifeartaigh models with general R-charge assignments can have
vacua where both supersymmetry and R-symmetry are spontaneously broken. Most of
these vacua are metastable because the potential shows a runaway behaviour. We
explain the relation between runaway directions and R-symmetry.
\PACS{{11.30.Qc}{}-{12.60.Jv}{}} 
} 
\maketitle
\section{Introduction}
The recent discovery of metastable supersymmetry-breaking vacua in simple theories like $\mathcal{N}=1$ SQCD \cite{iss1} has opened new and promising directions in model building. Metastable supersymmetry breaking appears to be generic in field theories with strongly coupled gauge sectors. Previous difficulties with the SUSY-breaking sector have been overcomed and natural and promising models of gauge and direct mediation which include the SUSY-breaking sector can be constructed. 

The effective theory for these models can often be described by a Wess-Zumino superpotential. As an example we can consider the Intriligator-Seiberg-Shih model, which is the low-energy theory of $\mathcal{N}=1$ SQCD with $N_c < N_f < \frac{3}{2} N_c$ in the Seiberg-dual picture. This model corresponds to a weakly-gauged O'Raifeartaigh model: SUSY is broken by the Nelson-Seiberg argument \cite{ns} near the origin of field space, while the gauge dynamics restores the usual supersymmetric vacua far away in field space, so the non-supersymmetric vacuum is metastable but parametrically long-lived. 

From this point of view O'Raifeartaigh models are interesting because they can be effective theories of strong-coupling gauge theories, as in the ISS model, and therefore can encode the SUSY-breaking dynamics. R-symmetry plays a crucial role in these models: in fact the Nelson-Seiberg argument states that the existence of an R-symmetry is a necessary condition for supersymmetry breaking. 

There is a problem with R-symmetry in this picture. An unbroken R-symmetry forbids gaugino masses, which is phenomenologically unacceptable. The simplest possibility is to add terms that break the R-symmetry explicitly, or to consider the fact that R-symmetry can be anomalous when weak gauge dynamics is taken into account. However the absence of an R-symmetry in the theory implies that the SUSY-breaking vacua are metastable, therefore there could be some tension between the requirement of long-lived vacua and the bounds on gaugino masses. 

Another interesting possibility is the spontaneous breaking of R-symmetry. This is not constrained by requirements of vacuum stability and can be a useful ingredient in supersymmetric model building. Note that also in this case small effects which breaks  R-symmetry explicitly are needed to solve the problems related of the existence of a massless Goldstone boson (R-axion) originating from spontaneous breaking of the R-symmetry; however R-symmetries are expected to be accidental features of the effective theory which describes SUSY breaking, therefore small operators which break R-symmetry are expected to appear in the full theory.

Spontaneous R-symmetry breaking has been investigated in models with gauge interactions in \cite{dm},\cite{iss2}, where the Coleman-Weinberg potential is determined by Yukawa and gauge interactions. In this paper we are interested in spontaneous R-symmetry breaking triggered only by the classical and perturbative dynamics of O'Raifeartaigh models, without any contribution from gauge interactions.

Most of the O'Raifeartaigh models featured in the literature have fields whose R-charges can be chosen to be either 2 or 0. As discussed by Shih, in these models R-symmetry cannot be spontaneously broken. In fact their classical dynamics give rise to flat directions of the potential parametrized by some of the fields of R-charge 2 and these flat directions are lifted by 1-loop correction leaving a single vacuum at the origin of field space. Instead spontaneous breaking often occurs in models with fields with $R\neq 2,0$ \cite{shih}. 

The simplest model which breaks R-symmetry spontaneously for some values of its parameters is:
\be{W=fX+\lambda X\phi_{(1)}\phi_{(-1)}+m_1 \phi_{(3)} \phi_{(-1)} +\frac{1}{2}m_2 \phi_{(1)}^2\label{worig}}
where $R(X)=2$ and $R(\phi_{(k)})=k$. The flat direction parametrized by $X$ is lifted by quantum corrections and R-symmetry is broken if the resulting vacuum has $\langle X \rangle \neq 0$.

In this contribution we consider a generalization of the model \eqref{worig} with generic R-charge assignment. The superpotential has the form
\be{W=fX+\frac{1}{2}(M^{ij}+N^{ij}X+Q^{ij}_aY_a)\phi_i \phi_j \label{wmore}}
where $M,N,Q_a$ are generic symmetric complex matrices. In section \ref{ssb} we show that the vacua of these models break R-symmetry for a wide range of parameters of the superpotential.

In section \ref{run} we show that many Wess-Zumino models containing fields with $R\neq 2,0$ have runaway directions, both supersymmetric and non-supersymmetric, and explain the relation between the runaway behaviour and the R-symmetry of the theory. Finally we discuss the examples of models with runaway directions which appeared recently in the literature.

\section{Spontaneous R-symmetry breaking}\label{ssb}
We consider the model \eqref{wmore} with the condition $\det M \neq 0$. We can show that this condition is sufficient for supersymmetry breaking. To show this we write the vacuum equations $F_{\varphi^a}^\dagger=\partial_a W=0$:
\begin{eqnarray}
f+\frac{1}{2}N^{ij}\phi_i\phi_j=0 \label{ex}\\
\frac{1}{2}Q^{ij}_a\phi_i\phi_j=0 \label{ey}\\
(M^{ij}+N^{ij}X)\phi_j=0\label{ea}
\end{eqnarray}
 If $\det(M+NX+Q_aY_a)\neq0$ the only solution for the linear system \eqref{ea} is $\phi_i=0$ which cannot satisfy \eqref{ex}. R-symmetry implies that $\det(M+NX+Q_aY_a)=\det M$, therefore the equations cannot be solved simultaneously and SUSY is spontaneously broken \cite{shih}.

For small $f$ the classical flat directions given by $\phi_i=0$ are local minima of the potential. These directions are parametrized by $X,Y_a$ and they are lifted by the 1-loop Coleman-Weinberg potential generated for these fields. The quadratic part of the 1-loop potential at the lowest nonzero order in $|\hat
{M}^{-2}f\hat{N}|$ reduces to \cite{mio}:
   \be{V_{quad}= \frac{f^2}{32\pi^2}\int_0^\infty dv \ v^3 \mathrm{Tr}\left[\left|\mathcal{M}_1(v)\right|^2-\left|\mathcal{M}_2(v)\right|^2\right]\label{tt}}
 with
\be{\mathcal{M}_1(v)=\sqrt{2}vD\left(\hat{N}D^2\hat{Y}\right)D}
\be{\mathcal{M}_2(v)=D\left(\hat{N}\hat{M}D^2\hat{Y}+\hat{Y}\hat{M}D^2\hat{N}\right)D}
\be{D=\frac{1}{\sqrt{v^2+\hat{M}^2}}}
where \be{\hat{M}=\begin{pmatrix} 0 & M^\dagger \\ M & 0\end{pmatrix},\ \hat{N}=\begin{pmatrix} 0 & N^\dagger \\ N & 0\end{pmatrix}}\be{\hat{Y}=\begin{pmatrix} 0 & (NX+Q^aY_a)^\dagger \\ NX+Q^aY_a & 0\end{pmatrix}}
The potential (\ref{tt}) contains the mass terms for $X,Y_a$ at the origin.  If the whole expression is negative for some choice of $(X,Y_a)$ then the classical vacuum $X=0,Y_a=0$ is unstable and there can be an R-symmetry breaking vacuum along one of these tachyonic directions. 

 Note that the two terms in equation (\ref{tt}) are generally of the same order, but their contributions to the masses of $X,Y_a$ have opposite signs: positive for the first term, negative for the second one. For models containing only fields with $R=2,0$ the trace of the second term is always zero because of R-symmetry constraints on the form of the matrices $M,N,Q_a$, therefore R-symmetry is unbroken in these models \cite{shih}. 
 
For models with general R-charge assignment the contribution of $\mathcal{M}_2$ is generally non-zero and $V_{quad}$ can be positive or negative depending on the parameters of the model. In the last case there is spontaneous R-symmetry breaking. It can be shown numerically and analitically that spontaneous breaking occurs for a wide range of parameters $M,N,Q_a$ \cite{shih},\cite{mio}.

All these results are obtained in the limit of small $f$. When $f$ is of order $|M|^2/|N|$ the flat directions are no more local minima of the potential and the physics of these models is determined by another interesting feature, which will be discussed in the next section.

\section{Runaway directions}\label{run}

The SUSY-breaking vacuum of the model \eqref{worig} is metastable because of the existence of a runaway direction \cite{shih}:
\be{\phi_{(1)}=-\frac{f}{\lambda \phi_{(-1)}},\ X=\frac{m_2 f}{\lambda^2 \phi_{(-1)}^2} ,\ \phi_{(3)}=\frac{m_2 f^2}{m_1\lambda^2\phi_{(-1)}^3},\ \phi_{(-1)}\rightarrow 0}
This runaway direction goes to a SUSY vacuum at infinity, therefore the vacuum at $\phi_i=0$ is not the ground state of the theory.

To understand the reason for this runaway, we note that the runaway direction can be seen as a rescaling of all the fields of the model, given by a complexified R-symmetry transformation:
\be{\varphi(\epsilon) =\epsilon^{-R(\varphi)} \varphi(0) \quad,\quad \epsilon \rightarrow 0\label{resc}} 

We will see that this feature is not restricted to the Shih model. Almost all O'Raifeartaigh models with generic R-charge assignment have runaway directions \cite{mio}. The same is true for many Wess-Zumino models with R-symmetry and generic R-charge assignment which break SUSY because of the Nelson-Seiberg argument.

We present an argument which shows why runaway directions are a common feature of these models. The first step is the classification of the vacuum equations (and the F-terms) according to their R-charge:
\begin{align} \partial_i W=0,\ R(\varphi_i)<2 && R>0  \\  \partial_i W=0,\ R(\varphi_i)=2 & & R=0  \\  \partial_i W=0,\ R(\varphi_i)>2 & & R<0  \end{align}

The Nelson-Seiberg argument states that it is not possible to solve all these equations at the same time because of R-symmetry, therefore SUSY is broken for any finite value of the fields. However, it can be possible to solve a subset of these equations.

 Let's look at two common possibilities:
 \begin{itemize}
 
 \item In some cases it is possible to solve all the equations with $R\geq 0$ (or $R \leq 0$). In this case it is sufficient to rescale all fields by a factor  $\epsilon^{-R(\varphi)}$ (or $\epsilon^{R(\varphi)}$) as in \eqref{resc}. The rescaled fields are also solutions of $R\geq 0$ (or $R \leq 0$) equations. Then we send $\epsilon \rightarrow 0$ to solve also the equations with $R<0$ (or $R>0$) and obtain a supersymmetric runaway vacuum: $ V  \rightarrow 0$  as $\epsilon \rightarrow 0$. The direction $\varphi_a(\epsilon)$ parametrized by $\epsilon$ is the runaway direction.
 
This case often occurs when there are a few equations with $R=0$.  This is the case of model \eqref{worig}. In that model the equations with $R\geq 0$ are
  \begin{align} f+\lambda\phi_{(1)}\phi_{(-1)} & =0 & &R=0 \label{e1} \\m_1\phi_{(3)}+\lambda X\phi_{(1)} & =0 && R=3 \label{e2}\\ m_2\phi_{(1)}+\lambda X\phi_{(-1)} & =0 & & R=1 \label{e3} \end{align}
  and we can easily solve (\ref{e2}),(\ref{e3}) in terms of $\phi_{(3)},\phi_{(1)}$ and then rescale all fields $\phi$ to solve (\ref{e1}). It can be shown that all models (\ref{wmore}) with no fields $Y_a$ fall into this case \cite{mio}. 
  
Of course also models which do not have the form (\ref{wmore}) can fall into this case. An example is Witten's runaway model:
\be{W=fX+\alpha X^2\phi}
In this model the only equation with $R\leq 0$ can be solved and there is a runaway direction given by $f+2\alpha X\phi=0,X\rightarrow 0$ which flows to a supersymmetric vacuum at infinity.

 \item It often happens that it is not possible to solve the equations with $R=0$. 
 In this case we need to consider the fields which minimize the potential \be{V_{R=0}=\sum_{R(\varphi_a)=2}|F_{\varphi_a}|^2=\sum_{R(\varphi_a)=2}|\partial_a W|^2}
 and check if the equations with $R>0$ or $R<0$ (but not both) can be solved by the same values of the fields. This turns out to be possible in many cases. If this is possible, then the rescaling \eqref{resc} by a factor $\epsilon^{-R(\varphi)}$ (or $\epsilon^{R(\varphi)}$) does not change this property. The resulting value of the potential is $V=\min V_{R=0}+V_{R<0}$ and  the limit $\epsilon \rightarrow 0$ correspond to $V\rightarrow V_\infty=\min V_{R=0}$ which is the absolute minimum of the potential by construction. 
 
 This explicit example is a small modification of the Shih model with another pseudomodulus $Y$:
 \be{W=fX+(\lambda X+\eta Y) \phi_{(1)}\phi_{(-1)}+m_1 \phi_{(3)} \phi_{(-1)} +\frac{1}{2}m_2 \phi_{(1)}^2\label{wy}}
 The runaway direction is obtained as explained above:
 \begin{align}&\phi_{(1)}=-\frac{f}{\lambda' \phi_{(-1)}} ,\ X+\frac{\eta}{\lambda}Y=\frac{m_2 f}{\lambda'^2 \phi_{(-1)}^2}  ,\ \nonumber \\ &\phi_{(3)}=\frac{m_2 f^2}{m_1\lambda'^2\phi_{(-1)}^3},\ \phi_{(-1)}\rightarrow 0  \\& \lambda'=(|\lambda|^2+|\eta|^2)/\bar{\lambda}\nonumber\end{align}
 and the potential $V_\infty$ is the minimum of $V_{R=0}=|f+\lambda \phi_{(1)}\phi_{(-1)}|^2+|\eta  \phi_{(1)}\phi_{(-1)}|^2$. Most of models (\ref{wmore}) with a large number of fields $Y_a$ have
 similar runaway directions.
 
 Note that if it is possible to solve both the equations with $R>0$ and $R<0$ then the rescaling \eqref{resc} does not change the potential and therefore corresponds to a dilatation along the flat directions. The usual O'Raifeartaigh models with $R=0,2$ are examples of this case.
 \end{itemize}

Most of O'Raifeartaigh models with generic R-charges realize one of the two possibilities described above.

There are some interesting examples of runaway directions in the literature. Some of them show different mechanism of spontaneous R-symmetry breaking. 

The model of Essig, Sinha and Torroba \cite{essig} contains two interacting gauge sectors, but it can be described by an effective superpotential:
\be{W=\Phi \mathrm{tr} M+  \mathrm{tr}q M \tilde{q}+\Phi  \mathrm{tr}P{\bar P}+\left(\det P{\bar P}\right)^{-\frac{1}{N'_c-N'_f}}}
This superpotential has an accidental R-symmetry and the form of the last term implies that the R-charge assignment is different from $R=0,2$. In fact there is a runaway direction which goes to a supersymmetric vacuum. However, Coleman-Weinberg potential generates a metastable minimum near the origin of the runaway direction and R-symmetry is broken in this vacuum.

The model of Cho and Park \cite{cp} contains the ISS model and other singlets with a superpotential which is reminescent of (\ref{worig}). It is basically an embedding of the Shih model in the ISS model and R-symmetry is broken as seen in the previous section.

Abel, Durnford, Jaeckel and Khoze \cite{abel} considered instead the addition of  a baryon term to the ISS model:
\be{W=\tilde{q}_{i\alpha} M_{ij}q^\alpha_j+\mu^2 M_{ii}+m\epsilon^{rs}\epsilon^{\alpha\beta}q_{r\alpha}{q}_{s\beta}}
The baryon term implies an R-charge assignment different from $R=0,2$.
This model contains many fields with $R=2$, so it has a runaway direction which goes to a non-SUSY minimum. The Coleman-Weiberg potential lifts this non-SUSY runaway direction and stabilize the minimum at a finite value of the fields, breaking R-symmetry spontaneously.

\section{Conclusions}
O'Raifeartaigh models with generic R-charge assignment are interesting for model building because of the possibility of spontaneous R-symmetry breaking, which makes easier to obtain gaugino masses of the correct size. 

The presence of runaway directions in these models is an interesting and common feature. It is related to the R-symmetry of the superpotential by the argument discussed in this paper. The metastability of vacua near the origin can also be explained as a remnant of the supersymmetric vacua which appear when explicitly R-symmetry breaking terms are added to the superpotential \cite{mio}. 

 The presence of runaway directions can also be useful because of the existence of other R-symmetry breaking vacua along these directions, as in the above examples \cite{essig},\cite{abel}. It would be interesting to understand if the mechanisms for spontaneous R-symmetry breaking implemented in  these examples could be a generic feature of realistic metastable SUSY-breaking theories.

%
%

\end{document}